# LEARNING AND TEACHING EINSTEIN'S

# THEORY OF SPECIAL RELATIVITY:

# STATE OF THE ART

Xabier Prado[1], José Manuel Domínguez[2], Iván Area[3], José Edelstein[4], Jorge Mira[5,*], Ángel Paredes[6]

[1]Departamento de Física y Química, I.E.S. Pedra da Auga. Ponteareas. Pontevedra.
https://orcid.org/0000-0001-9535-7499

[2]Departamento de Didácticas Aplicadas, Universidade de Santiago de Compostela.
https://orcid.org/0000-0002-5159-8968

[3]Departamento de Matemática Aplicada II, Universidade de Vigo.
http://orcid.org/0000-0003-0872-5017

[4]Departamento de Física de Partículas, Universidade de Santiago de Compostela.
https://orcid.org/0000-0002-7485-9286

[5]Departamento de Física Aplicada, Universidade de Santiago de Compostela.
http://orcid.org/0000-0002-6024-6294

[6]Departamento de Física Aplicada, Universidade de Vigo.
https://orcid.org/0000-0003-3207-1586

*Corresponding author: jorge.mira@usc.es

# ABSTRACT

This work analyzes the difficulties in learning and teaching Einstein's theory of special relativity. An extensive bibliographic review has been performed, considering articles published in the most relevant journals on science education, which were selected taking into account the following impact factors: JCR, SJR, IN-RECS and ICDS. The typical thinking of students and teachers is discussed pointing out that, occasionally, it does not befit the proper scientific perspective. Different educational proposals are examined and particular didactic implications are inferred. The conclusions of this inquiry constitute the basis of a proposal that relies on a Minkowskian geometrical formulation for teaching special relativity in upper secondary education.

KEYWORDS

Einstein's theory of special relativity. Minkowskian geometry. Thinking of students and teachers. Didactic proposals. Bibliographic revision.

# 1. INTRODUCTION

Relativity, due to its media impact, has been and will keep on being a major topic for science popularization. As a result, it is somehow considered as an attractive legend, which is neither understood nor adequately perceived.

Einstein's theory of special relativity presents specific difficulties for its teaching and learning. Determining the state of the art of the topic may be considered a central issue for science education. Accordingly, the goal of this contribution is to provide, in the light of previous research, a global perspective of the problems associated to this subject. The points of view of students and teachers and different pedagogic programs are taken into account.

The insights resulting from this scrutiny have substantiated a didactic proposal, based on a Minkowskian geometric formulation (Minkowski 2012, Naber 2012), for the teaching of special relativity. The scheme was implemented and investigated with students of upper secondary school (mostly sixteen and seventeen years old) of a town of the autonomous community of Galicia (Spain). This work can be classified within the following research lines: Development of scientific thinking: reasoning schemes and action schemes and Design, implementation and evaluation of educational proposals utilizing Minkowskian geometry (Prado Orbán 2012).

# 2. METHOD

The selection of the analyzed bibliography was carried out using the following databases and criteria:

## 2.1. Retrieval of information. Documentary sources

The search of relevant literature was performed by means of the databases displayed in Table 1. They are specialized in human and social sciences, and contain comprehensive information concerning journals and resources related to education.

## 2.2. Analysis of reliability and validity of the information. Selection criteria

The selection criteria have been established taking into consideration:

The goal of the revision it to achieve a global vision about the problem of teaching-learning special relativity from the point of view of students, teachers and didactic methods.

The reliability and validity of the articles, taking into account their methodological quality and whether they fulfill the sought criteria of scientific quality. With this goal, the evaluation indices of scientific publications described in Table 1 were used.

---

*Journal Citation Reports (**JCR**)*.- It provides information regarding scientific journal in the fields of applied and social sciences. It is a part of the Web of Science. It includes two categories: Sciences and Social Sciences, and it is organized chronologically. The Institute of Scientific Information (ISI) of Thomson Scientifics is in charge of analyzing the journals, computing impact factors and spreading the information.

*Scimago Journal Rank (**SJR**)*.- This portal analyzes the bibliometric indices of 16.000 journals included in Scopus. It is an open access platform, allocated for the evaluation of impact and scientific performance of journals and countries. It is developed by the Scimago research group.

*Impact factor of Social Sciences Spanish journal (Índice de impacto de las revistas españolas de Ciencias Sociales*, **IN-RECS**). Worked out by the EC3 (Evaluation of Science and Scientific Communication) research group of the University of Granada (Spain). During the last trimester of every year, it published a list of journals, authors and institutions ordered by their impact. Its publication has been suspended and it was last updated in 2011.

*Composite index of secondary diffusion (**ICDS**)*.- An indicator measuring the dissemination of journals in scientific databases. It is published using the Matriu d'Informació per la Identificació i Anàlisi de Revistes Científiques (MIAR) database, prepared by the Library Science and Documentation Department of the University of Barcelona (Spain). It contains information about the quality of journals on Social Sciences and Humanities.

*Educational Resources Information Center (**ERIC**)*.- It is a database sponsored by the U.S. Department of Education. It is the main source of referential bibliographical information in Science Education, including more than 1.2 million citations. It is formed by two sources: Current Index to Journals in Education (CIJE) and Resources in Education (RIE). They both cover 14.000 documents and index about 20.000 journal articles every year.

*Google Scholar* is a freely accessible web search engine that indexes the full text or metadata of scholarly literature across an array of publishing formats and disciplines. The Google Scholar index includes most peer-reviewed online academic journals and books, conference papers, theses and dissertations, preprints, abstracts, technical reports, and other scholarly literature.

Table 1. Databases and impact factors taken into account for the selection of information.

## 3. RESULTS AND DISCUSSION

In order to systematize the results, perform the analysis and formulate the conclusions and didactic implications, the available information has been categorized. It is necessary to point out, however, that the resulting categories were inferred from the reviewed research, even if they have been reformulated or grouped in order to extract the difficulties of learning and teaching found in the context of English-speaking and Spanish-speaking communities. The goal has been to design, apply and evaluate a teaching proposal (Prado Orbán & Domínguez Castiñeiras 2010), based on the geometric features of Minkowski space-time diagrams.

The proposal was conceived in order for the process of conceptual change (Posner, Strike, Hewson & Gertzog 1982) to take place in two steps: a first one in which preclassical notions are addressed in order to achieve the desired understanding from the perspective of Galilean relativity, and a second one in which that knowledge, correct from the classical[1] point of view, is questioned in order to grasp the concepts of Einstein's special relativity. Accordingly, the following categories to organize the analysis of information have been established:

- − Thinking of the students in relation to Galilean relativity and to Einstein's theory of special relativity.
- − Thinking of the teachers about Einstein's theory of special relativity.
- − Didactic proposals for the teaching of Einstein's theory of special relativity.

### 3.1. Thinking and learning difficulties of the student body in relation to Galilean relativity

The theory of relativity, already in its classical formulation (Galileo, Newton), introduces a mathematical notion relative to the measuring scales: the origin of coordinates. In the same way, the pre-relativistic perspective considers that there is a zero point for motion, namely the state of rest. In this case, the scientific progress took the opposite direction to the usual one since it came to dismiss the absolute character of that zero point without substituting it by a more fundamental one. This might be one of the inherent complications for understanding the theory since it seems that, instead of laying stronger foundations, it introduces a larger ambiguity. Hence the known expression "everything is relative" as a popular and incorrect definition of the theory of relativity.

---

[1] As usual in science didactics, the term "classical physics" is used in contrast to "modern physics", that would include Einstein's relativity and quantum mechanics. Notice the difference with its meaning in physics texts, where "classical" is typically used as an antonym of "quantum".

In this sense, the Galilean principle of relativity presents epistemological dilemmas that must be the subject of a specific treatment before dealing with Einstein's theory of relativity, like the attribution of an absolute character to rest (Hewson, 1982). In this respect, Villani & Pacca (1987) found that students have problems in accepting that measurements of length and time may depend on the reference frame and, at any rate, attribute an "absolute" character to the measurements performed in a reference frame at rest with respect to the observer, a fact that is connected to the previous ideas.

Galili & Kaplan (1997), inspecting the presentation of reference frames in textbooks, realize that they are seldom used to introduce Newton's laws and to apply relativity, as it would be desirable. Even the conservation of momentum or energy is typically considered for only one observer. Saltiel & Malgrane (1980), in the context of Newtonian reference frames, indicate that the velocity, the travelled distance and the trajectory of an object in motion are seen as frame-independent.

Steinberg, Brown & Clement (1990) point out that Newton struggled for twenty years to get rid of the notions of "impetus" and "centrifugal force", what can mostly explain the difficulties students find when addressing those concepts.

Tefft & Tefft (2007) consider necessary to treat specifically the energy-related relativistic effects in classical relativity, formulating the energy conservation principles in a more general manner, adapted for the posterior teaching of special relativity. Díaz, Herrera & Manjarrés (2009) noted the complications in giving a proper meaning to the shift in work and energy when considering different classical reference frames, irrespective of whether they are inertial or not.

Duit (1984) synthetized five key aspects in relation to the concept of energy: the notion of energy, the transformation, the transference, the conservation and the degradation, to which the mass-energy equivalence put forward by the theory of special relativity should be added. However, Driver & Warrington (1985) pointed out the difficulties students face to distinguish work and energy and Solomon (1983) figured out a wrong usage of the concept of energy in the processes of transference.

Concerning functional relations and the status of physical constants, Viennot (1982) revealed that the word "constant" is understood by students in relation to the notion of number, rather than as a function of several variables that keeps a fixed value.

## 3.2. Thinking and difficulties of student learning in relation to the theory of special relativity

The complexity attached to a number of concepts bound to the theory of special relativity stems, mostly, from the necessity of freshly interpreting some phenomena that, intuitively, are considered "simple". When the pedagogy of the theory is addressed, one might expect the existence or emergence of alternative ideas among the student body. Numerous authors have examined this problem since the second half of the twentieth century. We have classified it in the following categories: Space and time, continuity and measurement, causality and simultaneity, reference frames, energy, mass, and light.

### 3.2.1. Space and time

In relation to space and time, the theory of relativity questions the most intuitive concepts acquired by humans over the years. Undoubtedly, the acceptance of a theory that challenges deep-rooted conceptions within human perception faces a firm opposition, as Hewson & Thorley (1989, p. 548) argue: the idea of "absolute time" remains unquestionable, as what "flows equally for everyone and everywhere", in a very similar formulation as the one established by Newton himself in his axioms. This spotlights the didactic problem caused by such ideas since, precisely, one of the consequences of the theory of relativity is the vanishing of the notion of absolute time.

In the framework of the theory of general relativity, the space-time acquires a marked substantial character. It can be interesting to deal with this aspect, especially in relation to geometric or visual approaches.

Levrini (2002, p. 265), when addressing the educational problem of general relativity, presents an analysis of the approaches from the point of view of ideas of "space", some of which can be useful for the teaching of special relativity. For instance, a crucial question regarding the nature of space had been debated by Newton and Leibniz and later by Einstein, Minkowski and Poincaré: is space a physical object endowed with substantiality, or is it just a set of formal relations between objects or their positions?

The concept of relativity itself can cause confusion if it is not adequately explained. For instance, when introducing the twin paradox, a careless application of the principle of relativity would lead to the conclusion that the reference frame moving with the travelling twin is as valid as the one moving with the brother staying at Earth. Thereby the paradox, since it seems evident that both cannot be older than the other one when they meet again. The conundrum arises from thinking that, if everything is relative to the reference frame, one might always choose the comoving frame for an accelerated observer. Assume that A is a static observer and B an accelerated one. It is clear that both comoving frames are not inertial with respect to each other. However, could one consider the B-comoving frame as "inertial" and the A-comoving as the non-inertial one? This line of reasoning is flawed because space-time is absolute and not relative, and, therefore, the acceleration with respect to space-time itself is a physical reality. Hence, it is possible to determine which observer is the really inertial one. It is interesting, in this respect, that Minkowski expressed discomfort with the choice of the term "relativity" by Einstein to refer to a theory that, as he had shown (Minkowski, 2012), essentially deals with the geometric properties of an absolute space-time.

### 3.2.2. Continuity and measurement

When analyzing these concepts from a formal logic point of view, those related to the continuity of velocity and, consequently, space and time come into play.

Taylor (2001) discussed the ideas of Leibniz and Newton with respect to time. Newton envisaged a time in permanent flow, even prior to the creation of the Universe. In sharp contrast, Leibniz conceived time as a relation between events and, thus, it would not make sense in their absence or in the absence of the Universe where they take place. The space is accordingly conceived from the relations between objects and would therefore be pointless if only a single object existed in the Universe.

Hawking (1988), in his renowned science popularization book about time and the Universe, discusses the traditional notion of time and then introduces the concept of imaginary time, that relates Minkowskian to Euclidean geometries and is pivotal for understanding a number of phenomena. The same idea is customarily used in quantum field theory, the framework developed by Richard Feynman among others that blends quantum mechanics and special relativity. This shows that the apparently simple concept of time is in fact tied to deep subtleties that add to the aforementioned difficulties.

Doménech, Doménech, Cassasus & Bella (1985) make an analysis comparing relativistic spacetime with classical space and time. They present a list of the properties of the aforementioned magnitudes, like passivity, independence, completeness, continuity, homogeneity and isotropy. In the case of classical relativity, they add the hypotheses of time universality, invariance of spatial distances and the Euclidean character of space, whereas for the special relativistic case, those hypotheses are substituted by relativity of spatial and temporal intervals within a pseudo-Euclidean spacetime within the geometric model put forward by Minkowski.

As commented above, Einstein's theory of relativity challenges firmly established ideas of classical physics, including the concept of measurement itself (which, incidentally, was also questioned by quantum mechanics). Provided the geometric nature of our teaching proposal, we find interesting the work of Barrett, Clements, Klanderman, Pennissi & Polaki (2006, p. 189), that accentuates the necessity of achieving precise measurements before addressing concepts like the universality of lengths.

### 3.2.3. Causality and simultaneity

The concept of causality, also known as cause/effect relations, is essential for a correct understanding of the theory of relativity and its consequences. The principle might be stated in the following way: "if A causes B, B cannot cause A", namely, an effect cannot precede its cause.

However, the concept of "absolute time" is challenged by special relativity and intuitive notions like "simultaneity" break down. Events that take place at the same time in an inertial reference frame are not necessarily simultaneous in a different one.

If simultaneity becomes a relative notion, it is natural to wonder whether cause/effect relations lose their absolute character and whether cause and effect might be swapped when considering different reference frames. It turns out that this does not happen in special relativity and, thus, it constitutes a consistency check of the theory. As an aside, it is worth mentioning that this is a non-trivial issue for general relativity since Gödel proved that solutions including closed time-like curves exist, apparently opening the possibility of traveling backwards in time. In this case, it is necessary to resort to chronology protection conjectures to avoid the problem of the violation of the causality principle.

Moreover, the theory of special relativity introduces the notion of "causally disconnected events": since velocities are limited by the speed of light, what happens somewhere in spacetime cannot affect other events taking place at some other point of spacetime that cannot be reached by light, even if it is in the future. This is fundamental for the theory and has far-reaching consequences. For instance, it misled Einstein himself in challenging the standard interpretation of quantum mechanics since he famously argued that the collapse of the wavefunction upon measurement would induce a "spooky action at a distance". It was only decades after Einstein's death when it was demonstrated by the outstanding experiments of Aspect (among others) that the "spooky action at a distance" indeed happens, but without breaking the causality principle. This is a clear example of how the causal connection of events in special relativity theory entails conceptual subtleties that eventually become difficulties for proper student learning.

Scherr, Shaffer & Vokos (2001, p. 528-533) probed the understanding of the concepts of time and simultaneity in special relativity by university students of scientific and non-scientific degrees. They presented an account of typical errors regarding these concepts: the misguided belief that two events are simultaneous if an observer receives signals from both at the same time, the conviction that simultaneity is absolute and the assumption that every observer constitutes a different reference frame. Following this study, Scherr, Shaffer & Vokos (2002) proposed the following strategies to evaluate and sort out the alternative ideas:

- With respect to reference frames, the proposal is to guide students in the determination of the time of an event taking into account the instant where a signal is received from it, in the construction of a reference frame from the measurements, and in the definition of simultaneity in a given reference frame.

- Regarding the relativity of simultaneity, they recommend to guide students in the application of the principle of the invariance of the speed of light, identifying and acting upon the difficulties related to causality.

- Concerning the concept of simultaneity itself, they propose the evaluation of the understanding of the concept of reference frame, relativity of simultaneity and the skills to solve quantitative problems that require its comprehension.

### 3.2.4. Reference frames

McDermott (1993, p. 295 y 297), comparing the way teachers teach and students learn, observed the difficulties the latter have for relating the spacetime graphs with the motions they represent and the concepts of energy and momentum, and emphasized the necessity of developing these skills.

Hewson (1982) analyzed explanations of relativistic phenomena and came to the conclusion that students only deemed real what is at rest with respect to themselves. Measurements performed for a moving body – e.g. the lifetime of a particle or a length – fall short of the reality of those made for a body at rest: it seems length and time are different, the existence of an absolute reality associated to rest is internalized.

The cited authors conclude that, typically, the complexity of the classical notions is not appreciated. When later dealing with relativity, the notion of privileged reference frames persists, influencing the perception of what is real or not.

### 3.2.5. Energy

In the classroom, the interplay between energy exchange and reference frames is seldom considered, especially when dealing with kinetic energy or work (Galili & Kaplan, 1997). Diaz, Herrera & Manjarrés (2009) pointed out that work and energy differ from the point of view of disparate classical reference frames. Interestingly, they found that in the center of mass reference frame the conservation principle always holds without fictitious forces, irrespectively of whether the frame is inertial or not.

Fort (2005) presented an application of the change of reference frame in order to incorporate time in the energy relations, an approach that is usually ignored when solving dynamical problems by taking into account energy conservation. In this way, apart from broadening the applicability range of the energetic treatment of dynamical problems, the relativistic ideas are substantiated, gaining credibility. Tefft & Tefft (2007) emphasize the need of specifically discussing the relativistic effects on energy in the framework of classical relativity as a preliminary step to the teaching of special relativity.

In order to discuss the difference between kinetic energy and momentum, Riggs (2016) advocates for a presentation that minimizes the mathematical formalism, both for classical and special relativity. On the other hand, Dib (2013) and Serafin & Glazek (2017) discuss the contribution of bonding forces, associated to potential energy, to the total mass-energy budget of a bound system.

### 3.2.6. Mass

Students work with diverse notions in relation to mass, depending on the framework (Doménech, Casasús & Doménech, 1993): quantity of matter (associated to the notion of atoms), conservation principle (associated to chemistry), velocity-dependent mass (associated to science-fiction contexts), disappearance of mass in nuclear reactions, etc. A definition in terms of quantity of matter gives rise to difficulties for identifying mass in an immaterial system like a couple of photons moving away from each other (Parasnis, 1998).

The notion of relativistic mass, or velocity-dependent mass, brings about an added complication because of its contrast with the accepted notion of invariant or proper mass. The utilization of the concept of relativistic mass leads to confusing concepts such as transverse and longitudinal mass, which differ from each other for a given particle in a given reference frame. Moreover, the relativistic mass puts the focus on the particles, rather than on the intrinsic properties of spacetime, as it would be desirable (Whitaker, 1976).

Warren (1976) highlighted the wrong notions of disappearance and appearance of mass in nuclear reactions, whose origin lies on the confusion between the sum of the components of a system, that may change, and its total invariant mass, that remains a constant.

These concepts are at the core of a debate (Doménech, 1998) – whose nature is often more semantic that physical – regarding the concept of relativistic mass (or velocity-dependent mass) that physically corresponds to the total energy including the classical "mass" and the kinetic energy. Hecht (2009) argues that Einstein was never in favor of the concept of relativistic mass and preferred the concept of invariant mass, which amounts to the rest mass for a single particle and to the total energy in the center of mass frame of a system of particles.

Hobson (2005, 2006a) proposes to present Einstein's formula, $E = mc^2$, in the sense of considering that the mass is equivalent to the energy of the field that underlies every particle and, for systems of particles, to the overall thermal kinetic energy. Hecht (2011) analyzes the problems Einstein had to tackle for establishing the equation, what allows for anticipating those that students might find.

In order to address the difficulties that have been set forth above, our teaching proposal draws on an adequate geometric visualization of the theory of relativity. It involves the recognition of the mass as the modulus, in the Minkowskian sense, of the energy-momentum vector (p, E). The same geometric formulation leads to the visualization of the energy, and not just the mass, as the magnitude defining the inertial properties of a body. The definition works both for a particle and for a system of particles.

### 3.2.7. Light

The speed of light is a key element in the formulation of the theory of relativity, according to which it is one of the fundamental constants of nature. For this reason, many researchers deem the presentation of some experimental evidence of its value to be of great didactic value. For instance, Stauffer (1997) designed an experiment for secondary school students that allows for a measurement of the wavelength of the radiation of a microwave oven, or Gülmez (1997) used optical fibers of different lengths and an oscilloscope to determine the speed of light.

Analyzing the validity and universality of the relativistic phenomenon of a limiting velocity, together with the difficulties for understanding the associated concepts, is considered to be essential. In this sense, Salart, Baas, Branciard, Gisin & Zbinden (2008) performed experiments to check the physical existence of superluminal velocities. It is known that, for example, a shadow, a laser spot in the surface of the Moon or a photon in particular regions of the expanding Universe (Drummond & Hathrell, 1980) can move faster than light, but they do not imply physical violation of the limiting velocity principle. This limiting velocity idea is closely tight to the theory of relativity and it is considered virtually consubstantial to it. This may lead students to believe that experiments as those mentioned above prove that the theory of relativity is, in itself, relative and that the advance of science might overcome the speed limitation.

Villani y Pacca (1987) pointed out that, in issues related to the speed of light, the ideas of absolute motion and fluctuating notions between what is real and feigned prevail. This is true even among university students, and the following argument consistently comes up: two observers seem to measure different values whereas, actually, a single quantification of space, time and speed of light exists.

In this case, we again encounter the obstacle that students reason about complicated problems using arguments of logical and formal character, without resorting to spacetime diagrams where the situations could fit in.

## 3.3. Difficulties and thinking of educators for the teaching of special relativity

Up to this point, we have mainly focused in the problems students confront to understand Einstein's special theory of relativity. Nevertheless, it is also interesting to review the difficulties that, possibly, teachers face in relation to this theory. The goal is to check whether, in view of the complexity of the involved concepts, they may, just like students, work with alternative ideas that are unsound from a scientific perspective. Numerous authors addressed this issue during the second half of the twentieth century. In this context, Alemañ Berenguer (1997) studied the presence of wrong concepts within teachers' understanding of this theory and revealed that:

− The idea that Einstein's equation, $E = mc^2$, is only useful to transform units of mass into energy or vice versa persists. The square of the speed of light is a conversion factor, analogue to a density. This interpretation obscures the universality of the equivalence between mass and energy with respect to the particular character of the density relations.

− An accelerated body never reaches the speed of light because the relativistic mass grows with velocity. This imprecision is related to the previous one.

− In relativity, the concepts of reference frame, system of coordinates and observer are completely equivalent. Even if these concepts are closely related, the notion of reference frame is essential for the treatment of relativity, but not all reference frames need to contain an observer.

− Relativity was born as the result of the dissatisfaction caused by the negative results of Michelson and Morley in trying to detect the absolute motion with respect to the ether. Actually, the theory of relativity emerged as the consequence of deep discrepancies between classical mechanics and electromagnetism. In fact, Einstein asserted that, even if he knew about the Michelson-Morley experiment, it had little or no influence in the development of his theory of relativity (Pais, 2005).

− Relativity assumes classical electromagnetism or, more generally, a massless field theory and, therefore, it should be taught as an extension of Maxwell-Lorentz electrodynamics.

− The most salient feature of relativity is that the speed of light is independent of the velocities of the source and of the observer. In this respect, Einstein's initial formulation of the invariance of the speed of light only referred to the former, with the apparently innocuous consequence that light does not propagate through some ether, but through empty space. His stunning conclusions stemmed from this fact.

− Relativity is a theory that deals with processes in extreme conditions of speed or energy. Thus, it would be better to dispense with it due to its theoretical complications and its alienation from daily life. Its teaching should be restricted to specialized courses. Against this kind of argument, it should be pointed out that relativistic effects do have applications and practical and tangible results. We thus believe that this idea should be gradually abandoned.

Mermin (1994) discusses the importance that some teachers grant to Lorentz matricial transformation as an essential element for teaching the theory of relativity. He compares that with what would be its analogue: explaining the concept of spatial rotation by introducing orthogonal matrices as the starting point. He argues that no professor would follow that path.

Alemañ (1997) found inappropriate applications of the concepts of reference frame and event, even in the framework of classical kinematics.

Baierlein (2006) addressed two common errors (or *myths*, as he calls them) in the teaching of relativity. The first one is the presentation of Galiean transformation as a kind of low-velocity limit for Lorentz transformation, a procedure that can cause confusion. The second one is the existence of two different approaches to the invariability of the speed of light that, again, can provoke disorientation in the student body.

Pérez & Solbes (2006) examined several distortions that take place in the teaching of relativity, the most common ones being:

- Introducing the topic with the "crucial" Michelson-Morley experiment. This empiricist approach leaves out the problematic situation.

- Focusing the development of relativity in the brilliant creation of a single man, Einstein, ignoring the collective nature of scientific advance.

- Omitting or, conversely, exaggerating the shift in the physical paradigms.

Ambrosis & Levrini (2010) made an empirical study to determine how physics teachers approach innovation by performing case studies of how teachers analyze and appropriate a teaching proposal. They argue that, even if the known difficulties show up in the implementation process of the proposal, they can already be identified and addressed in initial stages, when teachers are approaching the proposal and designing their own paths to it.

Sezgin &Selçuk (2011) checked that pre-service teachers at different academic levels of a Turkish university have specific and considerable difficulties with concepts like proper time, time dilation, proper length, mass and relativistic density.

De Hosson, Kermen & Parizot (2010) found a deep lack of understanding of the concepts of reference frame and event within a group of future teachers of physics. Some students think that events can be simultaneous for an observer but not simultaneous for another observer, even when both are in the same reference frame. On the same line, De Hosson & Kermen (2012) found that misunderstandings of this kind have the particular consequence that some future teachers believe that the order in which two events are perceived determines the order of succession of when they happened.

Turgut, Gurbuz, Salar & Toman (2013) realized that most of the candidates to a position of physics teacher at a pre-degree level had not been introduced to the theory of special relativity or its associated concepts. The students had difficulties with the relativity of time and with reference frames. Most of the candidates did not recognize the speed of light as a limiting velocity that no other object could reach.

Yavas & Kizilcik (2016) tried to figure out the reason why pre-service physics teachers faced difficulties in relation with special relativity, even if they found the topic to be interesting. They claim that problems with the mathematical formalism, the determination of the reference frame and the transition from classical to relativistic physics hindered their learning process.

## 3.4. Didactic proposals for teaching the theory of special relativity.

Finally, in order to complete the discussion about the state of the art of the subject of learning and teaching Einstein's theory of special relativity, this section contains a review and an analysis of the didactic proposals that have been put forward in the literature. We have classified them in terms of the didactic aspect that is mainly emphasized, and defined the following categories: *Proposals of conceptual change. Proposals of dissemination. Animations. Visualization without computation. Proposals of geometrical nature. Proposals based on the geometry of Minkowskian spacetime.*

### 3.4.1. Proposals of conceptual change

In this section, we discuss the didactic schemes that focus on the difficulties concerning relativistic concepts and propose general strategies to achieve the conceptual switch.

Pérez & Solbes (2003) commented on some problems with the teaching of special relativity, underlining the difficulties of paradigm change (classical or Galilean relativity to special relativity) and propose to work during lower secondary school (12 to 16 years old) to develop a reliable Galilean intuition. Analyzing previous notions concerning space, time and mass, they indicate that many of them were prompted by mass media, especially television (Allchin, 2004; Meneses-Fernández & Martín Gutiérrez, 2015).

Arriassecq & Greca (2007, p. 71 a 79) studied how special relativity is presented in textbooks in Argentina. They established a number of conceptual categories and subcategories: *historical contextualization* (state of the art of physics at the time, the concept of ether), *epistemology of the origin of the theory* (the role of experiments, originality, experimental checks, applications), *impact on other fields* (science, philosophy, art), *concepts* (space, time, simultaneity, relative motion, paradoxes). They analyzed the concept of time as change, the contraposition between absolute and relative time and the distinction between an observer (that measures physical magnitudes) and a spectator (that receives light signals and constructs images of the events).

Villani & Arruda (1998) attach particular importance to the history of science within the teaching of relativity in relation to the conceptual shifts, and propose several structural lines to address it.

Gil & Solbes (1993) propose a constructivist educational model to introduce modern physics starting from classical physics. They suggest overcoming it by taking into account different interpretations, like the transformation of mass in energy and the scarce internalization of ideas by the students.

Although the literature on the convenience of introducing contemporary physics in secondary education as well as the need to update their curricula are not abundant, some of the researchers are favorable to their teaching at an early age (Aubrecht, 1989; Swinbank, 1992, Ostermann & Moreira, 2000).

### 3.4.2. Proposals of dissemination

This category encompasses the programs focused in the simplification of contents in order to address broad audiences (dissemination) or for students at early ages. The main motive underlying this approach is the consideration of the theory of relativity as necessary knowledge for all fields, and not just for people that may eventually work with concepts of modern physics in their professional activities.

From the perspective of scientific dissemination, Johansson, Kozma & Nilsson (2006) discuss several museum experiences for the experimental visualization of particular aspects of the theory of special relativity, such as muons or antimatter. They also include an explanation about Minkowski and his ideas.

Sherin, Tan, Fairweather, & Kortemeyer, (2017) resort to a model with a much lower speed of light to present relativistic effects in a "human scale", in the context of the visit to a planetarium. In a similar fashion, Kortemeyer, Fish, Hacker, Kienle, Kobylarek, Sigler, Wierenga, Cheu, Kim, Sherin, Sidhu & Tan (2013) present and explain Gamow's scenario of a world in which the speed of light is comparable to that of a bicycle.

Valentzas, Halkia & Scordoulis (2007, p. 357-351) analyze the presence and utility of "thought experiments" in textbooks and dissemination books on modern physics, describing several classifications of them as put forward by several authors. Moreover, eleven thought experiments, most of them devised by Einstein, are listed and commented, whereas their educational applications are discussed: the pursuit of a light ray, magnet and conductor, train, light emission, fluid bodies, elevator, rotating disk, Heisenberg's microscope, Schrödinger cat, Einstein-Podolsky-Rosen paradox, a box with light and a clock.

A recurrent aspect in disparate didactic proposals for the theory of relativity and, more generally, of modern physics is the possibility (sometimes described as necessity, interest or convenience) of lessening the age at which students come into contact with these notions. Stannard (1990, p. 133) found that students know about the existence and (sometimes fantastic) properties of black holes from different extracurricular sources. Despite this fact, modern physics does not show up in the corresponding syllabus, hindering the possibility of a correct interpretation of that information that may help in building up a coherent set of ideas. He proposes to introduce modern physics at an early age, by presenting a series of conceptual glimpses. In his particle physics course, Swinbank (1992) suggests that all students, including those uninterested in studying physics, should receive information about modern physics, and in particular about special relativity.

Valentzas & Halkia (2013) use "thought experiments" to present relativistic concepts to secondary school students, whereas Dimitriadi & Halkia (2012) conclude that the basic ideas of relativity are accessible to secondary school students even if the following difficulties persist: a) the lingering of the idea of an absolute reference frame, b) the identification of objects with fixed properties, and c) events take place independently of the perception of the observer. Pitts, Venville, Blair & Zadnik (2014) studied the effect of didactic materials focused on aspects of special and general relativity on kids of late primary school (10-11 years old). The students were able to understand certain concepts such as the curvature of space, while for others, such as gravity in the Moon, no relevant progress was observed. Egdall (2014) describes a course on relativity for adults and discourages the use of the notion of relativistic mass.

### 3.4.3. Animations

The theory of relativity is closely connected to the integration of the fundamental concepts of space and time. Thus, it possesses an intrinsically dynamical character, which can be efficiently presented by animated visualizations that intend to benefit from the didactic potential of online audiovisual resources. Since the resources are becoming increasingly available in the World Wide Web, this possibility progressively gains importance.

Alonso Sánchez & Soler Selva (2006b) published a didactic proposal for upper secondary school based on thirty-one computerized animations created with the so-called *Modellus* software.

Savage, Searle, & McCalman (2007) consider the teaching at the university level and use the *Real Time* software to generate diverse graphical outputs that help in visualizing complicated relativistic effects.

Angotti, Caldas, Delizoicov, Rüdinger & Pernambuco (1978) proposed an attitudinal approach to the teaching of special relativity, presenting it as an intellectual challenge that requires an adaptation of the student's mental schemes in order to accept new and different ideas. With this strategy in mind, they start describing the dynamical aspects of the theory, including the notion of variable mass, by utilizing movies and other types of audiovisual material. In their conclusions, they reported that some students present an ambiguous attitude, accepting to live with a number of uncertainties while, at the same time, they do not identify relativity with aspects or problems of reality.

Holton (1962) published a list of books and articles that address the issue of the teaching of relativity.

Carson (1998) explained how to use a computerized spreadsheet to teach the theory of special relativity, presenting and example in which the value of the speed of light is computed by comparing values of the electric and magnetic fields.

Belloni, Christian, & Dancy (2004) use a software named *Physlets* and emphasize the importance of starting the teaching of special relativity by looking into the concept of reference frame. The visualization procedure that they described is not based on Minkowskian geometry, but on a systematic application of the concept of ray of light to define different events and their relations.

Pérez & Solbes (2006) present a didactic proposal for the teaching of special relativity as a motivation for the learning of physics. They consider that the didactic efficacy of a visual method based on Minkowski spacetime is limited. Instead, they resort to alternative visual approaches with light clocks, applications of Pythagoras' theorem, visualization software (*applets*). Moreover, they underscore the relations between science, technology and society as a motivation for learning.

McGrath, Wegener, McIntyre, Savage & Williamson (2010) proposed to use didactic material based of virtual reality and to check the advances attained through a final exam.

Sherin, Cheua, Tan & Kortemeyerb (2016) present *Open Relativity*, an open source online virtual laboratory in which users can propose their own experiments to explore the effects of special relativity.

### 3.4.4. Visualization without computation

Relativistic concepts are typically presented by means of mathematical language, which can be more or less abstract depending on the level of the students. In secondary school, the algebraic expression for the gamma factor, the velocity-addition formula or the mass-energy equivalence can be used. In higher education, more intricate methods such as hyperbolic trigonometry or tensorial calculus are suitable and allow for more complex computations. Nevertheless, focusing on these formalisms may obscure the comprehension of the actual physical processes as a coherent set of related notions. In order to address this problem, different authors have put forward proposals with emphasis on visualization, associated to a reduction of the mathematical formalism. This procedure intends to avoid the association of the theory of relativity with cumbersome equations and tedious computations.

Fiore (2000) recommends to teach the theory of special relativity using geometrical formulas like the Pythagorean theorem, examples including the operation of television tubes or muons and portraying Einstein as an unsociable student, in order to generate sympathy for his work.

Ruby (2009) substitutes calculus by simple algebra in the description of relativistic phenomena, presenting a introductory physics course called "Non Calculus" (Huggins 2011), that starts presenting the principle of relativity as a concept of universal validity. From it, the physical concepts of the XX and XXI centuries are gradually developed. This avoids confining the topic of special relativity to a few chapters at the end of the course.

### 3.4.5. Proposals of geometrical nature

The theory of relativity deals with the properties of space, time, other magnitudes derived from them and their transformations between reference frames. Presenting those magnitudes with spacetime diagrams offers opportunities for the visual understanding of relativistic concepts and for the performance of direct computations on the diagrams themselves, by making geometrical operations.

Romain (1963) presented a geometrical approach to the relativistic paradoxes, using spacetime diagrams for their explanation, visualization and resolution.

Dray (1989) put forward an alternative description and visualization of the twin paradox, applying spacetime diagrams defined on a circular space (cylindrical diagrams).

Mermin (1997) simplifies the visual form of spacetime diagrams disposing of the space and time axes and leaving as the main reference the relativistically invariant speed of light, which is represented by the tilted lines that define the light cone. The invariance of spacetime surfaces is applied to the resolution of different problems in 1+1 dimensional diagrams. The same author (Mermin, 1998), using these diagrams, presents a visual justification of relativistic invariants, identifying them with the constant surface of light rectangles.

Gron & Elgaroy (2007) use Milne's comoving coordinates to visualize the expansion of the Universe governed by Friedmann equations.

Silagadze (2008) studies from a geometric standpoint the existence of different spacetime models that fulfill the classical requirements for space (homogeneity, isotropy) and time (causality), together with the relativity principle. This is done without resorting to the experimental evidence regarding the speed of light and yields a set of geometries (Silagadze 2008, p. 833). The Galilean-Newtonian one does not fully conform to the principle of causality, and the so-called Carrol (or co-Galilean) geometry emerges as an alternative to it. In it, the effects of variation of simultaneity prevail over those of variation of position, what would be relevant, for instance, when analyzing events between distant galaxies. In the same contribution, the relation between the Big Bang and the geometry of spacetime is also discussed (Silagadze 2008, p. 851), considering the possibility that the cosmic microwave background constitutes an opportunity to retrieve the Aristotelian dynamical model of an absolute reference frame.

Kocika (2012) proposes to use a geometric diagram to visualize Poincaré's formula for the one-dimensional relativistic addition of velocities, whereas Dray (2013) explains three examples of problems of special relativity that can be solved by three-dimensional geometric diagrams rather than by algebraic calculations. The same author (Dray, 2017) presented didactic geometrical approaches to the teaching of special and general relativity, analyzing their pros and cons when compared to more traditional methodologies.

### 3.4.6. Proposals based on the geometry of Minkowskian spacetime

Our teaching proposal is mostly based on Minkowski's geometrical formulation of the theory of special relativity. Thus, we regard with great interest the geometrical formulations of the topic that have been published in the past. They basically follow from the geometrical nature of spacetime and the didactic possibilities of visualization and understanding that they allow for.

Alemañ Berenguer & Pérez Selles (2001) assert that the whole theory of special relativity is contained in the Minkowski diagrams and analyze several difficulties and doubts that tend to come up upon their utilization. They propose a didactic construction of Lorentz transformation starting from Galileo transformation and adopting a comparative visual procedure. They also advocate a change of the educational programs at the secondary school level in order to properly address the theory of relativity.

Taylor & French (1983) analyze a relativistic paradox concerning length in accelerated reference frames. They resort to Minkowski's geometrical formulation and, contradicting a common assumption, they show that the use of spacetime diagrams can be profitable to visualize and understand trajectories apart from free uniform motions. Also treating accelerated motions in spacetime diagrams, Desloge & Philpott (1987) explain scenarios related to black holes including event horizons of the existence of regions of spacetime from which not even light can exit to reach external observers.

Saletan (1997) adapted Minkowski diagrams to momentum space in order to visualize and solve the kinematic of particle collisions.

With the goal of inspecting whether teaching based on Minkowski diagrams is appropriate for upper secondary school, Cayul & Arriassecq (2015) implemented in Argentina a stage of a sequence designed to address special relativity in this fashion. They conclude that working with "pen and paper" diagrams is rather complicated and too much time is needed for the constructions. For this reason, Arriassecq, Greca & Cayul (2017) worked out a second implementation in which Minkowski diagrams were analyzed with *applets*. They argue that they enable a better conceptualization through the realization of qualitative and quantitative estimations. They indicate that the most important difficulties in the activities involving these concepts were related to numeric computations and to the correct management of algebraic equations.

Our teaching recommendation for special relativity is deeply rooted in the utilization of spacetime diagrams as didactic visual tools, in order to build up geometric reasoning and attain quantitative results. We close this section by describing two contributions that have been useful in the development of our proposal.

Müller (2000) analyzes the effect that the existence of an ether wind might induce on the global positioning system (GPS). He shows that the performance of the system with an error of tenths of meters can be considered as evidence for the invariance of the speed of light since, otherwise, the annual variation would be of at least two kilometers. The graphical representation in spacetime diagrams of the GPS operation in one spatial direction coincides with the explicative diagrams for the simplified Michelson experiment presented as dissenting experience in our visual proposal (see Section 5.2.2). He also analyzes the possibility of presenting the relativistic effect on space in two stages: first order approximations (as in the GPS case, where the effect comes from the loss of simultaneity due to the tilt of the spatial axis) and second order approximations (as in the full version of the Michelson experiment, where spatial contraction also shows up).

Likewise, Dryzek & Singleton (2007) interpret an experiment of positron annihilation as evidence for the absence of ether wind. The process can again be visualized with the same spacetime diagram used in the simplified version of Michelson's experiment and, therefore, it provides another possibility for presenting an experiment that departs from classical relativity.

## 4. CONCLUSIONS

The discussion presented in the previous sections constitutes an essential preliminary step for the construction of our teaching proposal. Thus, we present the conclusions stemming from the review of the state of the art in such a way that they will allow us to establish the didactic implications that can guide us in the design of the appropriate sequence of activities for teaching and learning Einstein's relativity using Minkowski's geometry.

### 4.1. Thinking of the students

The thinking of the students has to cope with the counterintuitive features of the theory of relativity. Frequently, this leads to the opinion that it is knowledge only within reach of a reduced group of people with serious training in modern physics. This opinion is strengthened because of the following:

- The theory is usually presented making use of complex mathematical formalisms such as tensors, matrices, complex numbers, four-vectors, etc.

- Students have preexisting ideas, which can be based on intuition (notions of absolute space and time, nuclear energy identified with combustion), on ordinary vague comments (*everything is relative*) or even on pseudoscience, sometimes in relation with science fiction.

- The great difficulty students face to substitute the classical physics concepts by the relativistic ones. This complication is reinforced by the lack of concrete daily experiences that would help in giving credibility or supporting the efficiency of the new theory.

- The influence of mass media that, occasionally, employ scarcely rigorous terminology, sometimes leading to confusion and wrong ideas.

Nevertheless, special relativity is one of the most precise and thoroughly checked existing theories, and technologies relying on it have a major impact on daily life. We believe that an adequate didactic treatment would help in understanding and appreciating the theory. Moreover, it would contribute to scientific literacy among non-experts. With this in mind, we have classified the main difficulties that students face in its study, as analyzed in the previous section:

### 4.1.1. Space and time

Students have problems to conceptualize the continuity features of space and time that show up when dealing with the concept of limiting velocity:

- In the Galilean case, velocity can grow indefinitely without reaching an infinite value. This means that the limit is infinity or, simply, there is no limit.

- In special relativity, velocity can constantly grow but, in any case, it cannot exceed a particular value, that of the speed of light.

- Students also show a singular resistance to accepting that the measures of length and time can depend on the reference frame, as they attribute an *absolute* character to measurements performed in the reference frame at rest with respect to the observer. Manifestly, the concept of Newtonian time, flowing equally in all reference frames, persists.

### 4.1.2. Continuity and measurement

Students are not able to get from a numerical algorithmic treatment the totality of the properties of the geometrical continuum. This is especially true in limiting cases, demonstrating a clear evidence for the failure of the integration of geometry and arithmetic. Moreover, there are difficulties in relation to visual comparisons of lengths of segments with different orientations.

### 4.1.3. Causality and simultaneity

There is little epistemological compromise with cause/effect relations and with wrong ideas regarding simultaneity. Nothing can be asserted about it and an absolute or excessively relative character is attributed to it. A usual confusion is that between the simultaneity of two events and the reception of signals from them by the observer.

### 4.1.4. Reference frames

Observations from a reference frame at rest and from another one in motion are ascribed a different status of reality by students. An absolute character is assigned to rest, and confusions appear between observer and reference frame. Moreover, students struggle to identify spacetime graphs to the motion they stand for.

### 4.1.5. Energy

Students consider energy as a quasi-material entity and there is no awareness of the influence of the reference frame on kinetic energy and mechanical work. They have difficulties in admitting the shift of energy when the frame changes.

In our teaching proposal, a fundamental role is conferred to the center of mass frame because it allows for appreciating the symmetry of certain situations.

### 4.1.6. Mass

There is confusion between mass and quantity of matter, and mass is considered to change with velocity. The concept of relativistic mass provoked a controversy in the scientific sphere. The concept of invariant mass, equivalent to the total energy of a system in the center of mass frame is nowadays preferred. This concept is adequately visualized in Minkowski spacetime. The operational definition of inertial mass coincides with the previous definition in spacetime diagrams for an inelastic collision.

### *4.1.7. Light*

The limiting character of the speed of light is thought to be due to insufficiency of the actual technology, whereas the Galilean sum of velocities is maintained. There is confusion between the invariance of the speed of light and the feature of being a limiting velocity. Moreover, high speeds are thought to produce destructive effects on the objects. Mechanical properties are attributed to the speed of light in Einstein formula (that disappear when $c = 1$ units are used). News in the media regarding experiments with superluminal velocities in quantum mechanics are wrongly credited as refutations of the theory of special relativity.

## 4.2. Thinking of the teachers

Just like for students, the intrinsic complexity of some of the concepts related to the theory of special relativity brings about difficulties in the teachers' thinking. There still exist numerous alternative ideas to the adequate ones from a scientific perspective. In the reviewed bibliography, the following ones are presented:

- It is better to neglect the theory of relativity because it is plagued with theoretical complications and classical physics is enough to understand the world around us.
- Einstein's equation for the equivalence between mass and energy is only useful to transform units of mass into units of energy and vice versa.
- The mass is identified with relativistic mass, which grows with velocity (experience shows that the concept of relativistic mass generates confusion, being the invariant mass more appropriate from a didactic standpoint).

- In relativity, the concepts of reference frame, system of coordinates and observer are completely equivalent.

- Since relativistic effects concern our measurements of space and time, but not space and time themselves, there is no reason to change those concepts with respect to those of the physics before Einstein.

- Relativity assumes classical electromagnetism and therefore it should be taught as an extension of Maxwell-Lorentz electrodynamics.

## 4.3. Teaching proposals

While reviewing the bibliography, we have verified that there are a large quantity and variety of didactic proposals for special relativity, what is an indication of the relevance of the topic. As already mentioned, we have found convenient to group them in categories in order to construct the exposition: *Proposals of conceptual change. Proposals of dissemination. Animations. Visualization without computation. Proposals of geometrical nature. Proposals based on the geometry of Minkowskian spacetime.* The literature shows that using spacetime diagrams is an efficient procedure to answer and explaining questions, dilemmas and paradoxes of the theory. In order to address the teaching of the relativistic concepts, the proposals mostly include theoretical or experimental activities: audiovisual resources (sometimes available online), experiments and analogies, thought experiments, historical accounts, approaches starting from classical physics, use of geometry, etc.

Some researchers have suggested that the Minkowskian geometrical formulation is uninteresting for the teaching of relativity. Nevertheless, in the present work we have checked that the number of didactic proposals precisely based on the efficiency and visual potential of this approach is growing over the years.

In our case, applying and evaluating the didactic proposal in the classroom, we found that, against a common belief, Minkowski diagrams can be successfully applied to generic motions and not only to uniform straight ones. Moreover, they were useful in explaining phenomena associated to black holes such as event horizons and the existence of regions of spacetime that not even light could traverse to reach external observers.

# 5. IMPLICATIONS FOR TEACHING

## 5.1. General considerations

The theory of relativity deals with some of the most basic notions of physics including time, space or mass, and its consequences challenge common sense. Accordingly, its conclusions affect all the structure of physics, and a correct interpretation is essential to achieve an adequate comprehension of fundamental physical phenomena. It has far-reaching implications for the notion of time itself, the nature of light, electromagnetism, nuclear physics, particle physics and cosmology, among other fields.

The usual approach to the theory of relativity through mathematical formalisms gives rise to considerable difficulties for its correct application. This is the reason why most people know some aspects of relativistic phenomena in an incomplete, partial and deformed way. This fact hinders the acceptance of the theory as a coherent body of physical concepts. In this work, the difficulties that students of different educative levels face when getting started with Einstein's special relativity have been reviewed.

In order to introduce the theory of relativity, one could think of using an exclusively kinematic approach, directly leading to the notions of time, space and relative velocity. In the case of special relativity, the integration of the energetic question — mass-energy equivalence through the expression $E = mc^2$ (actually, Einstein refers to mass through one of its properties, *inertia*) — creates additional complications for students.

Shortly after Einstein presented the theory of special relativity, the German mathematician Hermann Minkowski showed that all its physical consequences were of geometric character and could be formulated in a four-dimensional spacetime (Minkowski, 2012, Naber, 2012). In this framework, the mass corresponds to the invariant modulus of a four-vector whose components are energy (that, classically, is a scalar) and momentum (that, classically, is three-dimensional vector). The integration of both magnitudes in a more general physical entity with geometric properties is analogous to the integration of space and time in spacetime.

Notwithstanding, this four-vectorial treatment of energy and mass creates undue complications for the learning of the theory from a dynamical point of view. Several authors have proposed to dispose of it and to introduce mass in a way similar to that of classical physics, thereby emerging the concept of *relativistic mass*.

## 5.2. Outline of a teaching proposal

In view of the discussion presented in the previous sections, we believe that it would be interesting to devise a full program for the teaching of special relativity in terms of geometrical reasonings in Minkowski spacetime. The approach could be complemented with other methods described above, such as animations or historical narrative, or even with the most traditional algebraic procedure at some points. A preliminary version of this proposal was presented in (Prado Orbán & Domínguez Castiñeiras, 2010). The model has been designed, applied and evaluated in a sequence of activities in the classroom with promising results (Prado Orbán & Domínguez Castiñeiras, 2010). We expect to report on a fully developed version of the didactic scheme in the near future

The proposal is based on the utilization of Minkowski's spacetime in a simplified way, reducing it to one temporal dimension and just one spatial dimension. This approach allows the student to visualize some results of the theory and is useful to construct a didactic model. With it, it is possible to explain qualitatively and quantitatively relativistic phenomena like time dilation, spatial contraction, the existence of a limiting velocity and even the mass-energy equivalence. This geometrical-visual treatment does not lack the necessary rigor and validity to describe coherently special relativity. It can be used in university (Callaghan, 2000) to present the full theory and even the main aspects of general relativity. In the following, we provide some evidence of the potential of the setup by presenting in detail two examples of typical relativistic phenomena that can be described in simple visual terms. The first one corresponds to a relativistic elastic collision (Prado, Area, Paredes, Domínguez Castiñeiras, Edelstein & Mira, 2018) whereas the second one deals with a simplified version of the Michelson-Morley setup.

### 5.2.1. A relativistic elastic collision

Figure 1 displays, in a (p, E) diagram, a head-to-head (one-dimensional) elastic collision of a body of mass $m_1 = 8$ versus another one with $m_2 = 12$. Natural units $c = 1$ are used. The initial momentum-energy vectors are $I_1 = (-6, 10)$ and $I_2 = (16, 20)$. The initial velocity for each body can be obtained from the equation $u = p/E$, from which it is immediate to get $u_1 = -3/5$, $u_2 = 4/5$.

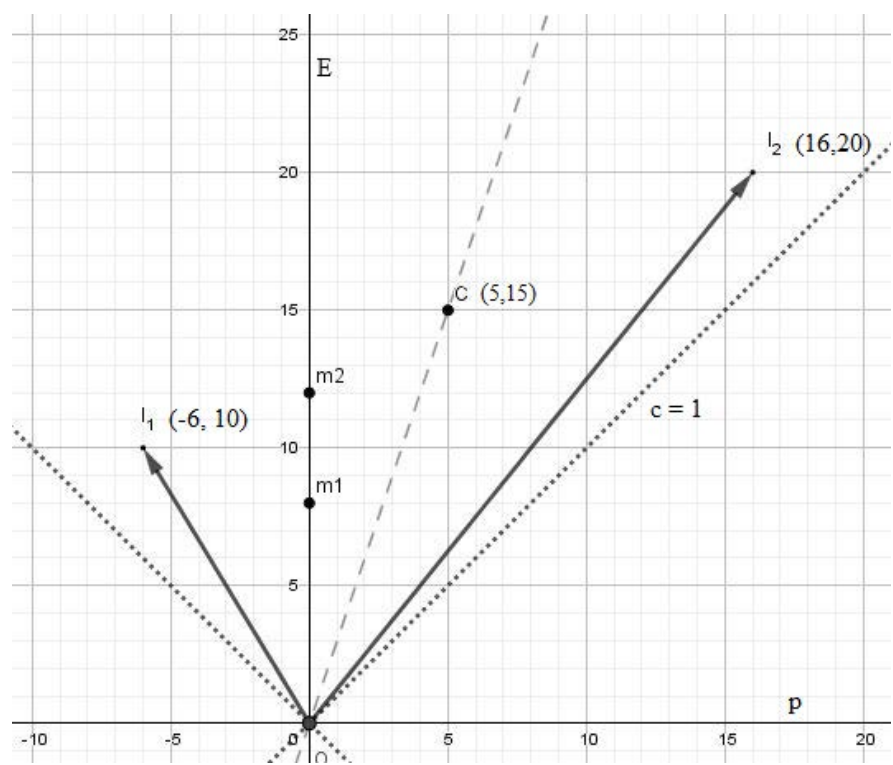

Figure 1- Frontal collision. The initial (p,E) values are {$I_1$, $I_2$}. C is the midpoint.

In the same figure, the midpoint between $I_1$ e $I_2$ is called C = (5,15). This point lies on the spacetime line for the center of mass, which moves with velocity $v_C$ = 5/15 = 1/3.

The kinematics of the elastic collision is solved graphically in figure 2. From $I_1$, $I_2$, we draw lines with slopes $v_C$ = 1/3 and $1/v_C$ = 3. The fact that these quantities are the inverse of each other is a direct consequence of applying Lorentz transformation to the collision described from the center of mass C reference frame. In this particular frame, the parallelogram corresponds to a rectangle whose sides are vertical (v = 0) and horizontal (v = 1 / 0 = infinity). We find a parallelogram, with vertices in the initial points ($I_1$, $I_2$) and the final points ($F_1$, $F_2$), that directly yields the solution to the collision.

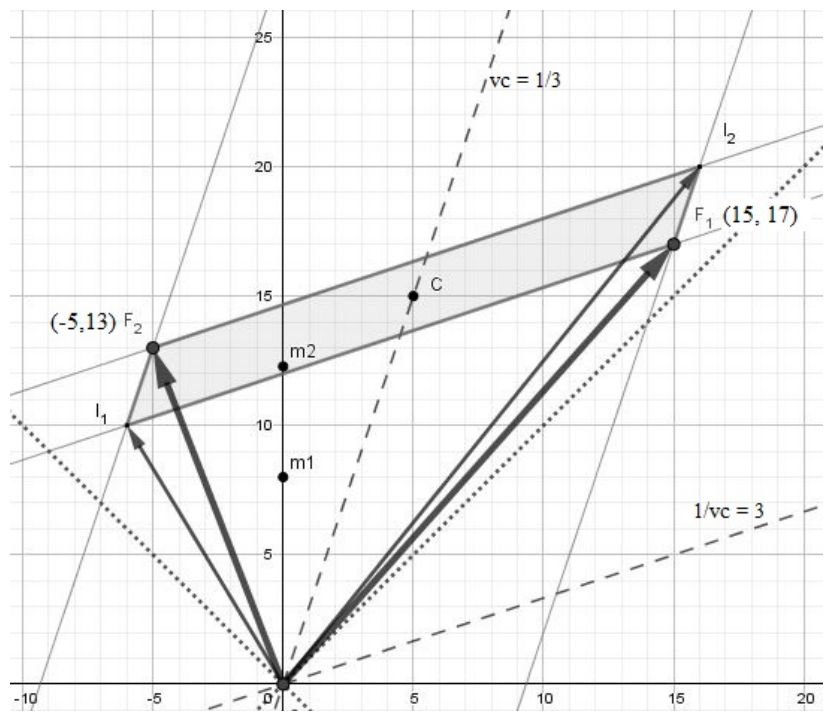

Figure 2- Parallelogram and final values ($F_1$, $F_2$)

The final values are found to be $F_1$ = (15,17) and $F_2$ = (-5,13) from which the final velocities are derived $v_1$ = 15/17, $v_2$ = -5/13. Thus, a single diagram integrates the geometric properties of spacetime (that acquire physical significance), the initial masses, energies and momenta and the final masses, energies and momenta, which directly solve the collision by simply constructing a parallelogram.

In order to verify the explicative efficiency of the geometric construction, we may compare it to the algebraic expression for the general solution of a relativistic one-dimensional elastic collision:

$$v_1 = \frac{2m_1 m_2 c^2 u_2 Z + 2m_2^2 c^2 u_2 - (m_1^2 + m_2^2)u_1 u_2^2 + (m_1^2 - m_2^2)c^2 u_1}{2m_1 m_2 c^2 Z - 2m_2^2 u_1 u_2 - (m_1^2 - m_2^2)u_2^2 + (m_1^2 + m_2^2)c^2}$$

$$v_2 = \frac{2m_1 m_2 c^2 u_1 Z + 2m_1^2 c^2 u_1 - (m_1^2 + m_2^2)u_1^2 u_2 + (m_2^2 - m_1^2)c^2 u_2}{2m_1 m_2 c^2 Z - 2m_1^2 u_1 u_2 - (m_2^2 - m_1^2)u_1^2 + (m_1^2 + m_2^2)c^2}$$

where $Z = \sqrt{(1 - u_1^2/c^2)(1 - u_2^2/c^2)}$.

Introducing the aforementioned values for the masses and velocities, we find Z = 12/25 and the values of the final velocities $v_1$ = 15/17, $v_2$ = -5/13, that agree with those found from the graph. Moreover, the relativistic kinematics is visualized while avoiding the contrived introduction of the relativistic mass, but without giving up the clarity of exposition that should help the students in acquiring a better significance of the phenomena.

### 5.2.2. Geometrical thinking for (a simplified version of) Michelson-Morley experiment

The necessity of finding an alternative to Galileo transformation was evident to Einstein from considerations concerning electromagnetism because Maxwell equations imply that the velocity of electromagnetic waves in vacuum is the same for all inertial reference frames. However, the mathematical complexities of this reasoning obscure its physical significance, and introducing special relativity in this way involves difficulties for the students. This is especially true when dealing with early ages and dissemination contexts.

In 1890, Michelson and Morley tried to measure the speed of the Earth travelling through some absolute space where a material substance named ether was supposed to be static. The idea was to detect differences in the speed of light travelling in different directions. Their null result was only understood several years later with the advent of special relativity. The Michelson-Morley experiment involves a comparison of light rays propagating in perpendicular directions to each other. This is not easily represented in the aforementioned didactical contexts. However, we have seen in section 3.4.6 that it is possible to discuss alternative experiences, which are equivalent in the sense of implying a necessity of postulating the invariance of the speed of light, stemming from experimental facts, e.g. (Müller, 2000) and (Dryzek and Singleton, 2007).

Let us know discuss how these results provide experimental evidence for a non-Galilean transformation. We start by indicating the features that a spacetime transformation should have in order to comply with experimental evidence. First of all, Newton's first law implies that the transformation should be linear since uniform motion in any fiducial inertial reference frame should transform into uniform motion in any other inertial frame. In spacetime, uniform motion is represented by a straight line, and therefore we conclude that straight lines have to be transformed to straight lines. Moreover, parallelism must also be preserved, because if two objects do not meet in a reference frame, they should also not meet in an alternative one. In order to implement a geometrical-visual representation of transformations between inertial reference frames, we can use two-dimensional diagrams (with time and one spatial dimension). Then, any square in the initial frame must become a parallelogram in the transformed one. Figure 3 depicts the transformation in classical physics, where time is assumed to be absolute.

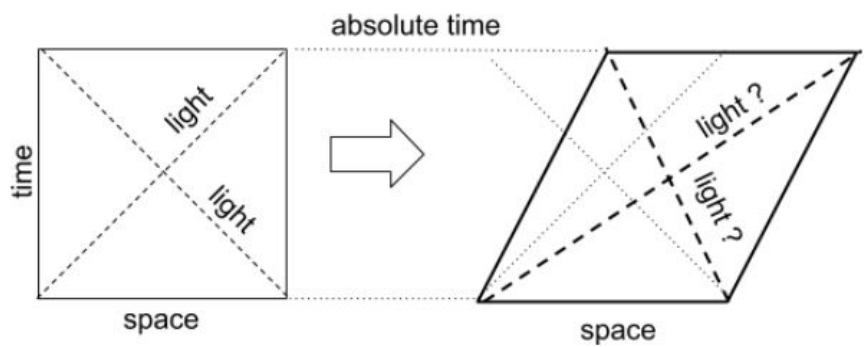

Figure 3- Galilean transformation. Natural units with speed of light equal to one (in the initial frame) are used.

Horizontal dotted lines joining events in both reference frames are a consequence of the absolute time hypothesis. From this pictorial representation of Galileo transformation, it is apparent that the dashed lines connecting the opposite vertices do not correspond to the speed of light of the initial reference frame. This experimental discrepancy leads to the introduction of an alternative transformation, in which the oblique lines joining the vertices of the parallelogram maintain their initial slope. Geometrically, this implies that the diagonals of the parallelogram must be perpendicular to each other. The only type of parallelogram with perpendicular diagonals is a rhombus, and it must have an inclination of 45º in order to preserve the slopes. Figure 4 depicts the solution.

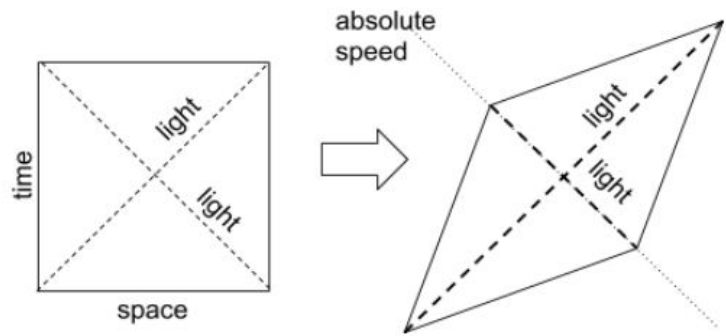

Figure 4- Lorentz transformation. Natural units with speed of light equal to one (in both frames) are used.

From the graph, one of the striking fundamental properties of Lorentz transformation becomes evident: the base gets tilted and initially simultaneous events lose that character in the transformed frame. The rest of relativistic effects can also be derived from the picture.

In order to univocally establish the form of Lorentz transformation, another property should also be taken into account: the conservation of areas within the spacetime diagram (Mermin, 1998). This requirement is immediately satisfied in Galilean transformations. In Lorentz transformation, it means that the stretching of a diagonal implies the (inversely proportional) dwindling of the other one. As a result, the area of the rhombus, equal to half of the product of the diagonal lengths, remains constant. From this geometrical feature, quantitative important aspects of special relativity can be derived, including the value of the gamma factor and the formulae for the relativistic Doppler effect.

# ACKNOWLEDGEMENTS

This work has been supported by Agencia Estatal de Investigación (AEI) of Spain under grant MTM2016-75140-P (co-financed by the European Union fund FEDER), by Ministerio de Economía y Competitividad (Spain) under grant EDU2015-66643-C2-2-P (corresponding to the project *Prácticas científicas en la enseñanza y aprendizaje de las ciencias. Dimensiones en la transferencia y el desempeño*) and grant FIS2017-83762, and Xunta de Galicia, grant R 2016/022.